\documentclass[twocolumn,showpacs,preprintnumbers,amsmath,amssymb,nofootinbib]{revtex4}

\usepackage{graphicx}
\usepackage{dcolumn}
\usepackage{bm}
\usepackage{amsmath}
\usepackage{psfrag}
\usepackage{color}
\usepackage{latexsym,amsfonts}
\usepackage{graphicx}
\providecommand{\href}[2]{#2}   
\definecolor{Blue2}{rgb}{0.,0.,0.8125}
\definecolor{Brown3}{rgb}{0.625,0.25,0.}
\definecolor{Cyan4}{rgb}{0.,0.56,0.56}
\definecolor{Green4}{rgb}{0.,0.56,0.}
\definecolor{LtBlue}{rgb}{0.27,0.42,0.52}
\definecolor{Magenta4}{rgb}{0.5625,0.,0.5625}
\definecolor{Red2}{rgb}{0.8125,0.,0.}
\newcommand{\be}{\begin{equation}}
\newcommand{\ee}{\end{equation}}
\newcommand{\bee}{\begin{eqnarray}}
\newcommand{\eee}{\end{eqnarray}}

\newcommand{\bs}{\begin{split}}
\newcommand{\ens}{\end{split}}

\begin{document}


\title{Electro-Magnetic Waves within a Model for Charged Solitons}
\author{Dmitry Borisyuk}
\altaffiliation{Bogolyubov Institute for Theoretical Physics, 03143, Kiev, Ukraine}

\author{Manfried Faber}
 \email{faber@kph.tuwien.ac.at}
 
\author{Alexander~Kobushkin}%
\altaffiliation[Permanent address: ]{Bogolyubov Institute for Theoretical Physics, 03143, Kiev, Ukraine
        and
        Physical and Technical National University of Ukraine KPI,  
        Prospect Pobedy 37, 03056 Kiev, Ukraine
}
 \email{kobushkin@bitp.kiev.ua}
\affiliation{%
Atominstitut der \"Osterreichischen Universit\"aten\\
        Technische Universit\"at Wien, Wiedner Hauptstr. 8-10\\
        A--1040 Vienna, Austria
}%

\begin{abstract}
We analyze the model of topological fermions (MTF), where charged fermions are treated as soliton solutions of the field equations. In the region far from the sources we find plane waves solutions with the properties of electro-magnetic waves.
\end{abstract}

\date{\today}

\maketitle

%
The intrinsic beauty of the Skyrme model and the well-known success of its application to short-range forces and strongly coupled particles make it worthwhile to extend its philosophy to the description of long-range forces and electrically coupled particles. The model of topological fermions (MTF) attempts to realize such an idea~\cite{Faber,FKob}.

The MTF field, $Q(x)$, is an SU(2) field  parameterized by 
\be
\label{I.2}
Q(x)=\cos\alpha(x) + i\vec \sigma\vec n(x)\sin\alpha(x),
\ee
where $\vec \sigma$ are the Pauli matrices. The fields $\alpha(x)$ and $\vec n(x)$ are functions of the Minkowski coordinates $x^\mu=(ct,x,y,z)$. The  $\vec n(x)$ field is a three dimensional vector in internal (``color'') space \footnote{We use the summation convention that any capital latin index that is repeated in a product is automatically summed on from 1 to 3. The arrows on variables in the internal ``color'' space indicate the set of 3 elements $\vec{q}=(q_1, q_2, q_3)$ or $\vec{\sigma}=(\sigma_1, \sigma_2, \sigma_3)$ and $\vec{q} \vec{\sigma}= q_K \sigma_K$. We use the wedge symbol $\wedge$ for the external product between color vectors $(\vec{q}\wedge\vec{\sigma})_A=\epsilon_{ABC}q_B \sigma_C$. For the components of vectors in physical space ${\mathbf x}=(x,y,z)$ we employ small latin indices, $i,j,k$ and a summation convention over doubled indices, e.g. $({\mathbf E} \times {\mathbf B})_i = \epsilon_{ijk} E_j B_k$. Further we use the metric $\eta = \mathrm{diag}(1,-1,-1,-1)$ in Minkowski space.}. It is constrained by the condition $\vec n^2(x)=1$ and defines a two-dimensional sphere which further on we call $\mathrm{S^2_{col}}$.

The Lagrangian of the MTF reads
\be\label{MTFlag}
\mathcal L=-\frac{\alpha_f\hbar c}{4\pi}\left(\frac14\vec R_{\mu\nu}\cdot\vec R^{\mu\nu}+\Lambda(q_0)\right),
\ee
where $\vec R^{\mu\nu}$ is curvature tensor
\be
\vec R^{\mu\nu}=
\vec\Gamma^\mu\wedge\vec\Gamma^\nu, \label{Rmunu}
\ee
with the connection
\be
\vec \Gamma^\mu=\frac1{2i}\mathrm{Tr}(\vec\sigma \partial^\mu QQ^\dag)\label{connection}
\ee
and the potential term is given by
\be\label{potential}
\Lambda(q_0)=\frac1{r_0^4}\left(\frac{\mathrm{Tr}Q}{2}\right)^{2m}=
\frac1{r_0^4}\cos^{2m}\alpha(x), \quad m=1,2,3,\dots
\ee
The model contains two parameters, the fine-structure constant, $\alpha_f=e^2_0/4\pi\varepsilon_0\hbar c\approx 1/137$, and a dimensional parameter $r_0$.

Note that ``the curvature term'' $-\frac14\vec R_{\mu\nu}\cdot\vec R^{\mu\nu}$ is proportional to the Skyrme term, but the so-called kinetic term does not enter the Lagrangian (\ref{MTFlag}) in order to allow for electromagnetic fields and forces \cite{FKob}.

Due to its Lagrangian the MTF has different properties than the Skyrme model at $r\to\infty$ \cite{Faber,FKob}. In the Skyrme model the chiral field $U$ approaches the trivial configuration, $U\to 1$. In the MTF the field configuration is determined by the potential minimum, i.e. $\alpha(x)=\frac{\pi}2$ at $r\to\infty$. As a result the $Q$ field becomes nontrivial,
\be\label{Q_infty}
Q(x)=i\vec \sigma \vec n(x) \qquad \text{at}\qquad r\to\infty.
\ee

The field $\alpha(x)$ describes the profile of a charged soliton with properties of an electron, whereas the field $\vec n(x)$ is related to the dual electromagnetic field strength \cite{Faber,FKob} by
\be\label{EMfield}
^{\ast} f_{\mu\nu}(x)=-\frac{e_0}{4\pi\varepsilon_0 c}[\partial_\mu \vec n(x)\wedge\partial_\nu \vec n(x)]\cdot \vec n(x).
\ee
The field strength $f_{\mu\nu}$ reads $f_{\mu\nu}=-\frac12\epsilon_{\mu\nu\rho\sigma}{^{\ast} f^{\rho\sigma}}$ with $\epsilon^{0123}=1$.

In the wave zone, where $\alpha(x)\to\pi/2$, the field $\vec n(x)$ should describe the free electromagnetic field. This can be shown by  solving the equations of motion in the wave zone and comparing them with the solutions of Maxwell's equations.

In the wave zone the equations of motion for the field $\vec n(x)$, derived in \cite{FKob}, are
\begin{equation}\label{EOM}
\partial^\mu\vec{n}\;\partial^\nu\left\{[ \partial_\mu \vec{n}(x) \wedge \partial_\nu \vec{n}(x) ]\cdot \vec{n}(x)\right\}=0.
\end{equation}
Due to the identity $\vec{n}\cdot\partial_\mu\vec{n}=0$ these are two independent equations only.

The aim of this paper is to solve Eqs.~(\ref{EOM}) and to show that these solutions behave like electromagnetic waves. 

In terms of the vector field $\vec n(x)$ electric $\mathbf{E}$ and magnetic $\mathbf{B}$ fields are defined by
\be\label{EandB}
\begin{split}
E_i&={\textstyle\frac12}\kappa\epsilon_{ijk}\left(\partial _j \vec n \wedge  \partial _k \vec n \right)\cdot\vec n,\\ 
c^2B_i&=\kappa \left(\partial _t \vec n \wedge  \partial _i  \vec n\right)\cdot\vec n,
\end{split}
\ee
where $\kappa =-e_0/(4\pi\varepsilon_0)$ in the International System of Units (SI) \cite{PDG}.

In the wave zone the electric and magnetic fields, propagating along the z-direction should satisfy the constraints \cite{Jackson}
\be
\label{3.3B}
\begin{split}
E_z&=\kappa \vec n \left( 
\partial _x\vec n \wedge  \partial _y \vec n \right)\cdot\vec n =0,\\
c^2B_z&=\kappa \left(\partial _t \vec n \wedge  \partial _z  \vec n\right)\cdot\vec n=0.
\end{split}
\ee

According to the constraint $\vec n^2=1$, the field $\vec n$ has two degrees of freedom. We describe these two degrees of freedom in terms of two variables, $\zeta(x^\mu)$ and $\eta(x^\mu)$. The constraints (\ref{3.3B}) can be fulfilled automatically by the following dependence of the $\vec n(x)$ field on $\zeta$, $\eta$ and the Minkowski coordinates 
\bee
\vec n = \vec n(\zeta(z,t), \eta(x,y,\zeta(z,t))).
\label{nform}
\eee
Below we show that the variable $\zeta(z,t)$ is a function of $z\pm ct$ only.

\begin{figure}[t]
\centering
\psfrag{x}{$x$}\psfrag{y}{$y$}\psfrag{e}{$\varepsilon$}
\psfrag{E}{$\mathbf E$}\psfrag{B}{$\mathbf B$}
\psfrag{Esin}{$E\sin \varepsilon$}
\psfrag{Ecos}{$-E\cos \varepsilon$}
\psfrag{Bsin}{$B\sin \varepsilon$}
\psfrag{Bcos}{$B\cos \varepsilon$}
\includegraphics[width=0.35\textwidth]{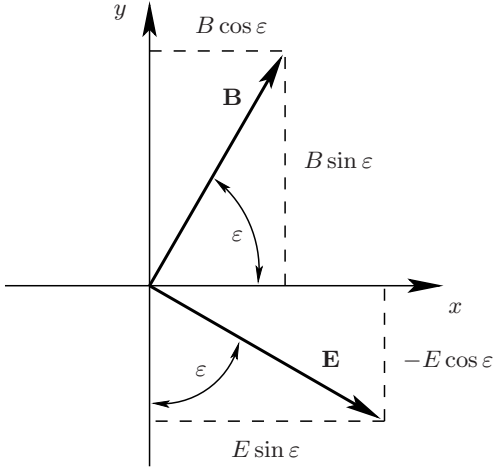}
\caption{The physical meaning of the angle $\varepsilon(\zeta(z,t))$.}
\label{polangle}
\end{figure}

In terms of $\zeta$ and $\eta$ and using (\ref{nform}) the electric and magnetic fields (\ref{EandB}) take the form
\begin{eqnarray}
\begin{aligned}
&\mathbf{E} 
  = \kappa \vec n \cdot (\partial_\zeta \vec n \wedge \partial_\eta \vec n) \;
      \partial_z\zeta
     \left(-\partial_y\eta,\,\partial_x\eta ,\, 0\right),\\
c^2&\mathbf{B}
  = \kappa \vec n \cdot (\partial_\zeta \vec n \wedge \partial_\eta \vec n) \;
     \partial_t\zeta
    \left(\partial_x\eta,\, \partial_y\eta,\ 0\right).
\end{aligned}
\label{Strength.1}
\end{eqnarray}
Introducing the notations
\begin{eqnarray}
&\cos\varepsilon=\displaystyle{\frac{\partial_x\eta}{|\nabla_\perp\eta|}},\quad 
\sin\varepsilon=\displaystyle{\frac{\partial_y\eta}{|\nabla_\perp\eta|}},\\
&\text{where} \quad \nabla_\perp\eta\equiv (\partial_x\eta,\,\partial_y\eta),
\nonumber
\label{polarization}
\end{eqnarray}
we get
\begin{eqnarray}
\begin{aligned}
&\mathbf{E} 
  = \kappa \vec n \cdot (\partial_\zeta \vec n \wedge \partial_\eta \vec n) \;
      \partial_z\zeta|\nabla_\perp\eta|
     \left(-\sin\varepsilon,\,\cos\varepsilon ,\, 0\right),\\
c^2&\mathbf{B}
  = \kappa \vec n \cdot (\partial_\zeta \vec n \wedge \partial_\eta \vec n) \;
     \partial_t\zeta|\nabla_\perp\eta|
    \left(\cos\varepsilon,\, \sin\varepsilon,\ 0\right).
\end{aligned}
\label{Strength.2}
\end{eqnarray}
The parameter $\varepsilon$ has the meaning of an angle between the $x$ axis and the magnetic field (Fig.~\ref{polangle}) \footnote{$\varepsilon$ together with the propagation direction defines the polarization plane as it was experimentally defined in crystal optics, ``Fresnel definition''.}. 

Using the spherical angles  $\theta$ and $\phi$ in color space
\be
n_x=\sin\theta \cos\phi,\quad n_y=\sin\theta \sin\phi,\quad n_z=\cos\theta
\label{sphCoord}
\ee
we can relate the factor $\vec n \cdot (\partial_\zeta \vec n \wedge \partial_\eta \vec n)$ in Eqs.~(\ref{Strength.1}) to the ratio of area elements between the two sets of internal coordinates $(\cos \theta, \phi)$ and $(\eta,\zeta)$. We will use an area preserving mapping from $(\eta,\zeta)$ to $(\cos \theta, \phi)$ and get
\be
\vec n \cdot (\partial_\zeta \vec n \wedge \partial_\eta \vec n)
 =\frac{\partial(\cos\theta,\phi)}{\partial(\eta,\zeta)} = 1.
\label{mapping}
\ee
For a special solution we will show later that this condition can result in topological restrictions on possible types of  waves.

Of course such a mapping is not unique. This means that there may be different realizations for the non-observable $\vec n(x)$ field, which leads to the same physical fields $\mathbf{E}$ and $\mathbf{B}$.

Assuming that the coordinates $(\eta,\zeta)$ fulfill the condition (\ref{mapping}) one arrives at
\be\label{Strength.3}
\bs
&\mathbf{E} =
  \kappa \, \partial_z \zeta \,|\nabla_\perp\eta| (-\sin\varepsilon,\, \cos\varepsilon,\, 0),\\
c^2&\mathbf{B} =
 \kappa \partial_t \zeta\,|\nabla_\perp\eta| (\cos\varepsilon,\, \sin\varepsilon,\, 0\,).
\end{split}
\ee
The field strengths obviously fulfill the general condition for electromagnetic waves \cite{Jackson} $\mathbf{E}\cdot\mathbf{B}=0$. In order to identify the electric and magnetic fields in Eq.~(\ref{Strength.3}) with those in electromagnetic waves they should satisfy another constraint, $|\mathbf{E}|=c|\mathbf{B}|$. According to Eq.~(\ref{Strength.3}) this leads to the relation $|\partial_z \zeta|=|\partial_t \zeta|/c$. This constraint is obviously fulfilled if $ \zeta$ depends on $z_\pm=z\pm ct$. In order to show this we have to turn to the equations of motion (\ref{EOM}). Substituting (\ref{nform}) in the equations of motion (\ref{EOM}) one arrives at two coupled non-linear equations
%
\be
\begin{split}
 &\partial_\zeta
   \left(\nabla_\perp  \eta\right)^2
  \left[\left(\partial_z\zeta\right)^2 
  - \frac{1}{c^2}\left(\partial_t\zeta\right)^2\right]+\\
  &\hspace{1.5cm} +2\,\left(\nabla_\perp  \eta\right)^2
  \left[\partial_z^2 \zeta -\frac1{c^2}\partial_t^2 \zeta\right] = 0,
\\
 &\Delta_\perp \eta
  \left[\left(\partial_z\zeta\right)^2 
  - \frac{1}{c^2}\left(\partial_t\zeta\right)^2\right] =0, \\ 
&\Delta_\perp \eta\equiv (\partial^2_x+\partial^2_y)\eta.
\end{split}
\label{e-o-m.3.b}
\ee
%
These equations can be fulfilled by the solutions of
\be
\begin{cases}
\partial_z^2 \zeta -\frac1{c^2}\partial_t^2 \zeta = 0,\\
\left(\partial_z\zeta\right)^2 
  - \frac{1}{c^2}\left(\partial_t\zeta\right)^2 = 0.
\end{cases}
\label{TwoWaveEqu}
\ee

The first equation of the system (\ref{TwoWaveEqu}) is a usual wave equation with two partial solutions
\be
\zeta(z,t) = \zeta_+(z_+),\quad\text{or}\quad \zeta(z,t) = \zeta_-(z_-)\,,
\ee
where $\zeta_+$ and $\zeta_-$ are arbitrary functions.  The second equation in (\ref{TwoWaveEqu}) is a non-linear one. The partial solutions  $\zeta_+$ and $\zeta_-$, but not a superposition of them, satisfy this equation too. Thus we have shown that the constraint $|\partial_z \zeta|=|\partial_t \zeta|/c$ is fulfilled.

Hence, the electric and magnetic fields, given by Eq.~(\ref{Strength.3}), describe polarized ``electromagnetic waves'' with amplitudes $\kappa \, \partial_z \zeta \,|\nabla_\perp\eta|$  depending on the space-time point. This dependence is a consequence of the non-linearity of the equations of motion.

For the special case
\be
\label{specialcase}
\partial_\zeta \left(\nabla_\perp \eta\right)^2=0\quad\text{and}\quad \Delta_\perp \eta=0
\ee
the system (\ref{e-o-m.3.b}) is reduced to one equation, which is the wave equation $\partial_z^2 \zeta -\frac1{c^2}\partial_t^2 \zeta = 0$. The general solution for $\zeta$ is a superposition of two waves
\be
\zeta(z,t) = \zeta_+(z_+) + \zeta_-(z_-)\,.
\label{solutionzeta1}
\ee

Now let us discuss a special solution of Eqs.~(\ref{specialcase})
\be
\label{specsolution}
\begin{split}
&\eta=a(\zeta)x+b(\zeta)y,\\
&a(\zeta)=d^{-1}\cos\varepsilon(\zeta),\quad b(\zeta)=d^{-1}\sin\varepsilon(\zeta),
\end{split}
\ee 
where $d$ is some constant. It results in waves, which are independent of the $x$ and $y$ coordinates, like plane waves in Maxwell electrodynamics.

Putting
\be\label{EundB}
\varepsilon=\mathrm{const}\quad \text{and}\quad \zeta=\cos kz_\pm
\ee
one gets linear polarized electromagnetic waves
\be
\bs
\mathbf{E} &= \kappa\frac{k}{d}\sin kz_\pm (\sin \varepsilon,-\cos \varepsilon,0 )\,, \\
c\mathbf{B} &=\mp \kappa\frac{k}{d}\sin kz_\pm (\cos \varepsilon,\sin \varepsilon,0 )\,,
\end{split}
\label{linpolwaves}
\ee
where $k$ is the wave number. Now we have to show that a mapping to the coordinates $(\cos\theta,\phi)$ exists, which satisfies the condition (\ref{mapping}). Such a mapping is
\be
\cos\theta = \zeta, \qquad \phi =-\eta.
\label{linpolmap}
\ee

\begin{figure}[t]
\centering
\psfrag{x}{$x$}\psfrag{y}{$y$}\psfrag{e}{$\varepsilon$}
\psfrag{a}{$d$}
\includegraphics[width=0.35\textwidth]{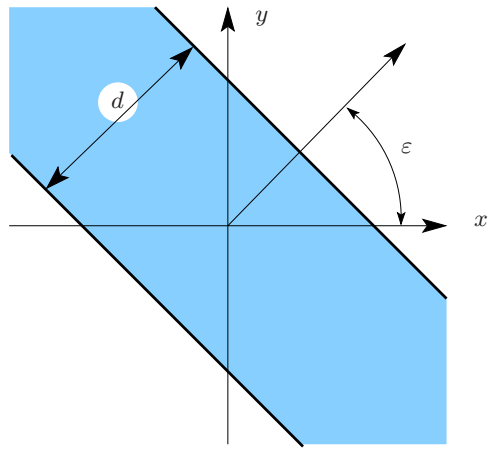}
\caption{Region in the $(x,y)$-plane, where circular polarized waves exist.}
\label{xyregion}
\end{figure}

For circular polarized waves the situation is not so simple. In this case the polarization angle $\varepsilon$ is a linear function of $z_\pm$, $\varepsilon=z_\pm k$, and the absolute value of the electric and magnetic fields is constant. With the choice
\be
\begin{split}
&\cos\theta = \eta=\frac1d(x\cos\varepsilon+y\sin\varepsilon),\\
&\phi = \zeta =  z_\pm/l,\qquad \varepsilon = k z_\pm,
\end{split}
\label{cirmap}
\ee
where $d$ and $l$ are arbitrary length parameters, we arrive at the required behavior of $\mathbf{E}$ and $\mathbf{B}$ 
\be
\begin{aligned}
\mathbf{E} &= \kappa /(dl) (-\sin kz_\pm,\cos kz_\pm,0 ),  \\
c\mathbf{B} &= \pm\kappa /(dl) (\cos kz_\pm,\sin kz_\pm,0 ).
\end{aligned}
\label{circ}
\ee
From (\ref{cirmap}) follows that the region of $\eta$ is restricted to
\be
\label{restriction}
-1\le\eta\le+1.
\ee
This means that the wave exists only on a strip of width $d$ in the $(x,y)$-plane parallel to the $\mathbf E$ field (filled region in Fig.~\ref{xyregion}). For given field-strength $E_0$ the width $d$ can be chosen arbitrarily large.

Let us investigate the topological necessity for the restriction (\ref{restriction}) in more detail. We have a similiar problem as mapping a globe onto a flat surface. In our case the globe corresponds to $\mathrm{S_{col}^2}$, the surface is a two-dimensional area perpendicular to the electric flux lines (\ref{circ}). Such an area is the helicoidal area
\be
x\sin kz_\pm-y\cos kz_\pm=0\quad\textrm{at}\quad t=0,
\ee
it has the topology of $\mathrm R^2$. The electric field strength is given by the ratio of an infinitesimal area on $\mathrm{S_{col}^2}$ to the corresponding area on the helicoid, see Eq.~(\ref{EandB}) and \cite{Faber,FKob}. Requiring constant field strength on the helicoid the mapping has to be area preserving. It is well-known that an area preserving mapping of a globe onto a plain is only possible in a restricted region, e.g. that defined in Eq.~(\ref{restriction}), $-1\le\cos\theta=\eta\le+1$.

For linear polarized waves the electric field strength is not constant on the  $z,r_\perp$-plane, it oscillates with $z$ and the mapping (\ref{linpolmap}) can be defined in the whole space.

Now let us discuss the connection of the equations of motion (\ref{EOM}) with electrodynamics. 
Introducing the abbreviation,
\begin{equation}\label{magcur}
\begin{aligned}
g^\mu&=\kappa\,\partial_\nu \hspace{0.2mm}\left\{[ \partial_\mu \vec{n}(x) \wedge \partial_\nu \vec{n}(x) ]\cdot \vec{n}(x)\right\}=\\
&=(c\rho_{\rm mag},\mathbf{g})=(c\nabla\cdot \mathbf{B},- \nabla \times \mathbf{E} - \partial_t\mathbf{B})
\end{aligned}
\end{equation}
the equations of motion (\ref{EOM}) reduce to
\begin{equation}\label{EOM_mag}
\partial_\mu\vec{n}\;g^\mu=0.
\end{equation}
The quantity $g_\mu$ is obviously conserved, $\partial_\mu g^\mu = 0$, and looks formally like a magnetic current \cite{Dirac}. But there is the essential difference to Dirac's magnetic current that $g_\mu$ has no external source, it is a result of the non-Abelian nature of the color field $\vec n(x)$.

Evidently, the equations of motion (\ref{EOM_mag}) are fulfilled by solutions of the homogeneous Maxwell equations, $g^\mu=0$. But the inverse does not hold true. In the wave zone the MTF equations of motion (\ref{EOM_mag}) substitute the homogeneous Maxwell equations.

According to Eqs.~(\ref{sphCoord}) and (\ref{mapping}) the magnetic current reads in terms of the parameters $\eta$ and $\zeta$
\be\label{mag.current}
\bs
& \rho_\mathrm{mag} 
  =\frac{\kappa}{c^2}\,\partial_t\zeta\,\Delta_\perp \eta,\\
& g_x = \kappa\left\{\frac{\partial^2\eta}{\partial x\partial \zeta}
      \left[\left(\partial_z \zeta\right)^2 
      -\frac{1}{c^2}\left(\partial_t \zeta\right)^2\right]+\right.\\
&\left.\hspace{2.cm}      + \partial_x \eta
        \left(\partial_z^2 \zeta -\frac1{c^2}\partial_t^2 \zeta \right)\right\},\\
& g_y = \kappa\left\{\frac{\partial^2\eta}{\partial y\partial \zeta}
      \left[\left(\partial_z \zeta\right)^2 
      -\frac{1}{c^2}\left(\partial_t \zeta\right)^2\right]+\right.\\
&\left.\hspace{2.cm}      + \partial_y \eta
        \left(\partial_z^2 \zeta -\frac1{c^2}\partial_t^2 \zeta \right)\right\},\\
& g_z 
  =-\kappa\,\partial_z \zeta\,\Delta_\perp \eta.
\end{split}
\ee
If there is no superposition of two waves $\zeta_+$ and $\zeta_-$\,, the $x$ and $y$ components of the magnetic current vanish, $g_x=g_y=0$. This condition is fulfilled for linear and circular polarized waves, (\ref{linpolwaves}) and  (\ref{circ}), respectively. In turn, if the condition (\ref{specsolution}) is fulfilled, $\rho_{\mathrm{mag}}$ and $g_z$ are also vanishing. The solution for linear polarized waves agrees with this condition at any point of physical space. But for circular polarized waves it is not so at the boundary of the strip of Fig.~\ref{xyregion} and $\rho_{\mathrm{mag}}$ and $g_z$ appear at this region. 

In summary, we discuss solutions of the model of topological fermions in the wave zone. The equations of motion are non-linear field equations. We find plane wave solutions of these equations with the properties of electromagnetic waves. We describe our solutions for linear and circular polarized waves and give topological reasons, why it is not possible to define circular polarized waves with a modulus of the electric and magnetic field strength constant everywhere in space-time.

\acknowledgments
The authors would like to thank Andrei Ivanov for numerous helpful discussions. This work was supported in part by Fonds zur F\"orderung der Wissenschaftlichen Forschung P16910-N12.

\end{document}